\documentclass[aps,prl,floatfix,twocolumn,superscriptaddress,showpacs]{revtex4}
\usepackage{amssymb, amsmath, graphicx}
\usepackage[usenames]{color}
\definecolor{Red}{rgb}{1,0,0}
\definecolor{Green}{rgb}{0,1,0}
\definecolor{Blue}{rgb}{0,0,1}

\begin{document}

%\preprint{}

\title{Characteristic Energy of the Coulomb Interactions and the Pileup of States}


%\author{}

\author{Daniel Mazur}
\email[Corresponding author: ]{mazur@anl.gov}
%\thanks{}
\affiliation{Materials Science Division, Argonne National Laboratory, Argonne IL 60439}
\affiliation{Physics Division, Illinois Institute of Technology, Chicago, IL 60616}

\author{K.~E.~Gray}
\affiliation{Materials Science Division, Argonne National Laboratory, Argonne IL 60439}

\author{J.~F.~Zasadzinski}
\affiliation{Physics Division, Illinois Institute of Technology, Chicago, IL 60616}

\author{L.~Ozyuzer}
\affiliation{Materials Science Division, Argonne National Laboratory, Argonne IL 60439}
\affiliation{Department of Physics, Izmir Institute of Technology, Izmir, Turkey}

\author{I.~S.~Beloborodov}
\affiliation{Materials Science Division, Argonne National Laboratory, Argonne IL 60439}
\affiliation{James Franck Institute, University of Chicago, Chicago, IL 60637}

\author{H.~Zheng}
\affiliation{Materials Science Division, Argonne National Laboratory, Argonne IL 60439}

\author{J.~F.~Mitchell}
\affiliation{Materials Science Division, Argonne National Laboratory, Argonne IL 60439}

\date{\today}

\begin{abstract}
Tunneling data on $\mathrm{La_{1.28}Sr_{1.72}Mn_2O_7}$ crystals confirm Coulomb interaction effects through the~$\sqrt{\mathrm{E}}$ dependence of the density of states.  Importantly, the data and analysis at high energy,~E, show a pileup of states: most of the states removed from near the Fermi level are found between $\sim40$~and 130~meV, from which we infer the possibility of universal behavior. The agreement of our tunneling data with recent photoemission results further confirms our analysis.
\end{abstract}


\maketitle

\begin{figure}
  \begin{center}
  \includegraphics*[width=3.3in, bb=34 235 590 560]{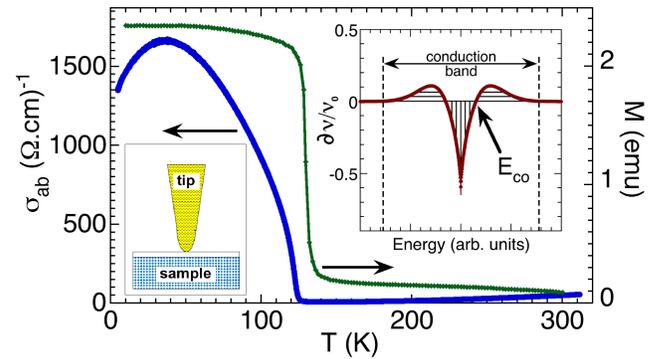}
  \caption{\label{fig:1}(Color online) Bulk $ab$-plane conductivity and magnetization (in a field of 1000 Oe) versus temperature are shown, arrows assign the scales to the curves. Right inset: Cartoon illustrating the expected Colomb interactions effect in 3D tunneling DOS. Equal areas with vertical and horizontal lines the state conservation, graph centered at Fermi energy. Left inset: The junction geometry.}
  \end{center}
\end{figure}

Coulomb interactions (CI's) between electrons can play a dominant role in strongly correlated systems and metal-to-insulation transitions. These include Mott insulators, complex oxides\cite{Higashiya07,Rienks07,Sarma98}, colossal magnetoresistive materials\cite{Mitra03}, and other novel metallic materials like graphene\cite{DasSarma07}. Coulomb interactions have also been studied as a possible mechanism of the pseudogap\cite{Emery95, Maly96, Millis06} in the density-of-states (DOS) of high-temperature superconductors. The Coulomb repulsion pushes some states near Fermi level, $\mu_F$, to higher energies, as described by Al'tshuler and Aronov (AA) \cite{Altshuler85}. Despite years of study, however, the essential implication of the theory, the conservation of states, has never been shown. In superconductivity most of the states depleted from $\mu_F$ pile up just above the energy scale gap, $\Delta$, but a few extend out as far as several times the Debye energy. The only upper limit for CI's is the band edge, but one may anticipate that most states will pile up just above some characteristic energy scale $\mathrm{E_{co}}$ (see right-hand inset of Fig.~\ref{fig:1}). 

The AA theory predicts the energy dependence for the DOS of correlated 3D metals to be $\sim$$\sqrt{E}$ for low energy, E, measured relative to $\mu_F$, and this dependence was confirmed by several tunneling studies\cite{Mitra03,Sarma98,Dynes81,McMillan81}. Photoemission spectroscopy (PES) showed a depressed spectral weight in metallic perovskites\cite{Sarma98} near $\mu_F$, and recently the $\sqrt{E}$ dependence was observed\cite{Kobayashi07} in PES spectra of $\mathrm{Sr_2FeMoO_6}$. Our measurements on bilayer manganite single crystals, $\mathrm{La_{2-2x}Sr_{1+2x}Mn_2O_7}$ (LSMO) for x=0.36, complement all these studies by also addressing conservation of states, a fundamental feature of electron-electron correlations, which has not been addressed in the literature to date.

Certain properties of LSMO make it ideal to study CI's. Low-temperature LSMO conductivity data (see Fig.~\ref{fig:1}) exhibit a $\mathrm{\sqrt{T}}$ dependence that is consistent\cite{Okuda99} with AA theory. The bilayered LSMO is a strongly correlated system, which is readily cleaved along the crystal $ab$-plane. This enabled detailed Fermi surface studies using angle-resolved photoemission (ARPES), which show\cite{Sun06,Mannella05} renormalized Fermi velocities, $\mathrm{v_F}$, of $\sim$1$-$3$\times$$10^7$~cm/s and electron scattering times, $\tau$, of $\sim$5$-$7~fs. Such small values define a poor metal, and one might expect them to be due to CI's in the DOS, $\mathrm{\nu(\mu_F)}$.

The effects of electron correlations on $\mathrm{\nu(E)}$, are \emph{directly} observable in the tunneling conductance, G(V)$\sim$$\mathrm{\nu(E=eV)}$$\propto$$\sqrt{V}$. A systematic analysis of the tunneling data yielded the values of $\mathrm{E_{co}}$, and an observation of a pileup of states within 2$-$3 times a characteristic energy $\mathrm{E_{co}}$ in analogy with superconductivity. Extracted values of $\mathrm{v_F}$ and $\tau$ agree with the published ARPES results, and they are consistent with an intuitive model that estimates $\mathrm{E_{co}}$$\sim$$\hbar/\tau$, effectively verifying that model.

Our tunneling data were taken on cleaved crystals of $\mathrm{La_{1.28}Sr_{1.72}Mn_2O_7}$ (i.e., x=0.36) at low temperatures, where they are metallic. Crystals were melt-grown\cite{Mitchell97} in an optical image furnace. The LSMO~(x=0.36) crystals were characterised  by measuring the temperature-dependent magnetization, $\mathrm{M(T)}$, as displayed in Fig.~\ref{fig:1} (conductivity data are from a different crystal). Crystals ($\sim$1$\times$1$\times$0.5~mm$^3$) were cleaved in air along the easy $ab$-plane, immediately mounted in the cryostat and cooled down to 4.2~K. Then the gold tip was brought into contact with the cleaved $c$-axis-normal crystal face to create junctions. Details of the apparatus are described elsewhere\cite{Ozyuzer98a}. 

\begin{figure}
  \begin{center}  
  \includegraphics*[width=3.1in, bb=145 110 490 625]{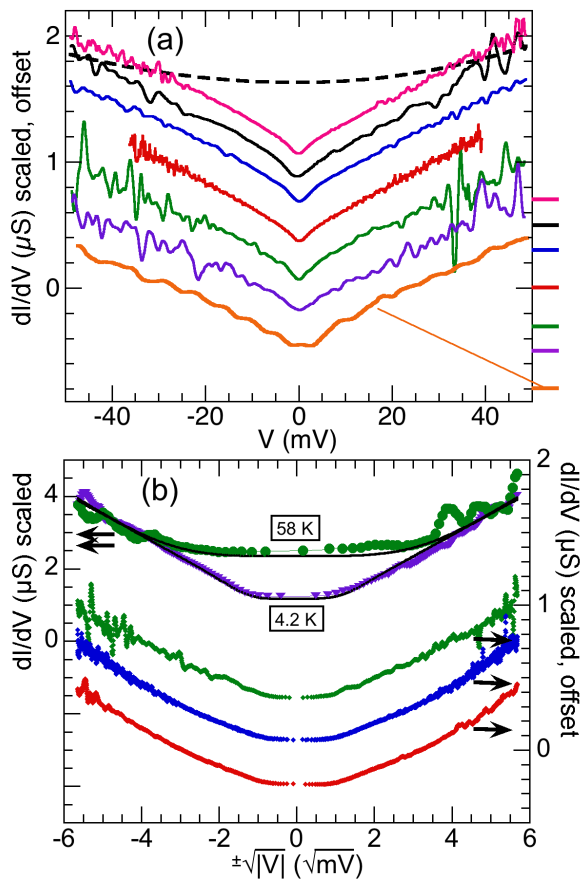}
  \caption{\label{fig:2}(Color online) (a) Set of G(V) curves at 4.2~K, topmost dataset overlayed with its $\mathrm{G_0(V)}$ fit. Zeroes for the curves indicated by bars on the right, in the same sequence. Zero-bias resistances of all junctions from bottom up were, in M$\Omega$: 80, 23, 2.6, 2.6, 1.8, 0.78, 0.46. The middle dataset is unscaled (values in $\mu$S). (b) Tunneling G(V) curves plotted versus $\mathrm{\sqrt{V}}$. Bottom 3 curves: The 4.2~K data to show nearly linear dependence. Upper 2 curves: A 58~K curve scaled on top of a 4.2~K curve (large datapoints). The $\mathrm{\sqrt{V}}$ feature is substantially thermally smeared in the 58~K curve, as predicted by the theory (solid black lines for 4.2 and 58~K calculated from Eqs.~(29) and (30) of Ref.~\cite{Abrikosov00}). The good agreement of model and data reinforces our conclusion that the zero-bias anomaly originates from electron-electron interactions\cite{fnote1}.}
  \end{center}
\end{figure}

For coherent tunneling, the reduced density of states, $\mathrm{\nu(E)/\nu_0}$ is $\sim$G(V)/G$_0$(V) in the low temperature limit, where G$_0$(V) is the barrier conductance that is independent of interaction effects and is approximately constant for barrier height, $\Phi$$>>$V. In Fig.~\ref{fig:2}a we present low-bias, G(V)$=$dI/dV, for several junctions measured at 4.2~K. The data are scaled and offset (bars on the right show zeros for each curve) to show the excellent reproducibility of the zero-bias cusp feature across two orders of magnitude of junction resistances. These curves have the approximate shape expected from the DOS effect of electron-electron interactions, $\mathrm{\nu/\nu_0(E)\propto\sqrt{E}}$, as can be seen in the plot of Fig.~\ref{fig:2}b versus $\mathrm{\sqrt{V}}$ that shows an extended linear region from $\sim$2$-$10~mV. However, our G(V) data represent the DOS convoluted with the barrier conductance G$_0$(V), which must be determined. A parabolic approximation to G$_0$(V) (see procedure below) is shown for the topmost G(V) dataset of Fig.~\ref{fig:2}a.

\begin{figure}
  \begin{center}  
  \includegraphics*[width=3.45in, bb=20 178 605 630]{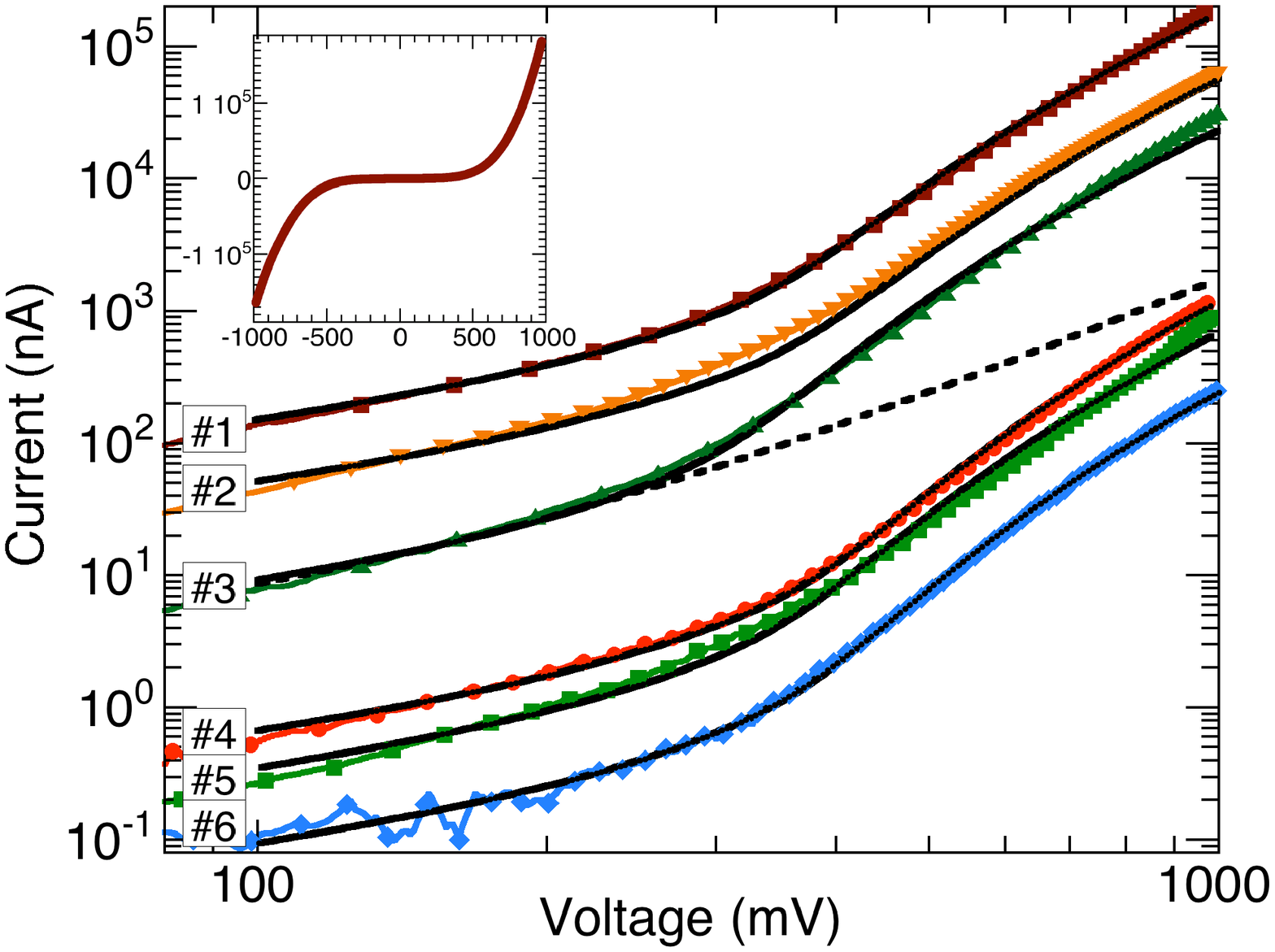}  
  \caption{\label{fig:3}(Color online) Demonstration of tunneling: Six unscaled tunneling I(V) curves (datapoints) on a log-log scale, with square-barrier WKB fits (solid black lines) of coherent tunneling across a square barrier. Resulting fit parameters paired as $\mathrm{t_0[nm],\Phi[meV]}$: \mbox{\#1: 1.55, 302;} \mbox{\#2: 1.53; 325;} \mbox{\#3: 1.73, 325;} \mbox{\#4: 1.64, 330;} \mbox{\#5: 1.67; 308;} \mbox{\#6: 1.73, 325}. Dashed line is a low-voltage $\mathrm{I_{fit}}$ curve fitting the WKB model \#3. Inset: Linear plot of the top-most I(V) curve from the main plot in the same units. }
  \end{center}
\end{figure}

Comparing our I(V) data with the WKB approximation allows us to prove that tunneling is their origin, to determine the barrier height and width, and to justify the parabolic approximation to G$_0$(V). Freeland et al.\cite{Freeland05} discovered that a non-ferromagnetic, insulating bilayer occurs naturally on the surface of LSMO (x$=$0.40) crystals and the extended I(V)'s shown in Fig.~\ref{fig:3} for x$=$0.36 closely follow the earlier data. The reasonable agreement over four orders-of-magnitude of current and three orders-of-magnitude of nominal junction resistance shown in Fig.~\ref{fig:3} confirms that tunneling is the primary conductance mechanism. Using the collected results of over 50 junctions we determine that $\Phi$=280$-$380~meV and t$_0$=1.1$-$1.8~nm\cite{fnote4}. The values of $\mathrm{t_0}$ reasonably approximate the full first surface bilayer thickness. Importantly, the excellent agreement with WKB tunneling model implies a high quality, uniform tunnel barrier.

%For the de-convolution of $\mathrm{\nu(E)/\nu_0}$ from the raw data, knowledge of G$_0$(V) is vital.  The transmission through a relatively low barrier (in our case $\sim$300 meV) is non-linear and causes strong V-dep\-en\-dence of G$_0$(V).   More recent data on LSMO~(x=0.36) show a similar non-magnetic surface bilayer and our data reproduce the former tunneling data qualitatively, see Fig.~\ref{fig:3}. Again we find very similar characteristics in junctions, whose variation in resistance over three orders-of-magnitude is likely due to different (unknown) junction areas.

%These data were successfully fit with a WKB approximation\cite{Holm35, Wolf85} of coherent tunneling across a~square barrier, shown as solid lines in Fig.~\ref{fig:3}. Square barrier is described by a height, $\Phi$, and thickness, t$_0$. Fits were made over the interval $\sim$100$-$1000~mV and, together with $\sim$50 more junctions, converged with parameter values t$_0$=1.1$-$1.8~nm and $\Phi$=280$-$380~meV\cite{fnote4}.  

However, the data over the full range of V necessarily include the higher-voltage effects of barrier rounding by image forces, barrier changes due to electrostriction, etc. that cannot be captured by our simple two-parameter WKB calculation. Thus we approximate the WKB model at V$<$$\Phi$ with a parabolic correction to a constant conductance, that is a current-voltage form $\mathrm{I_{fit}(V)=a(V}$$-$$\mathrm{V_0)+b(V}$$-$$\mathrm{V_0)^3+aV_0}$+$\mathrm{bV^3_0}$, where V$_0$ is the voltage shift of the parabola to account for a slight barrier asymmetry\cite{Brinkman70}. This approximation is shown by the dashed line in Fig.~\ref{fig:3} and is within 3\% precision for V$\leq$0.7$\Phi/e$, imposing a justifiable upper limit of $\sim$230~mV for use of $\mathrm{I_{fit}}$~\cite{fnote2}.

\begin{figure}
  \begin{center}
  \includegraphics*[width=2.9in, bb=165 130 455 625]{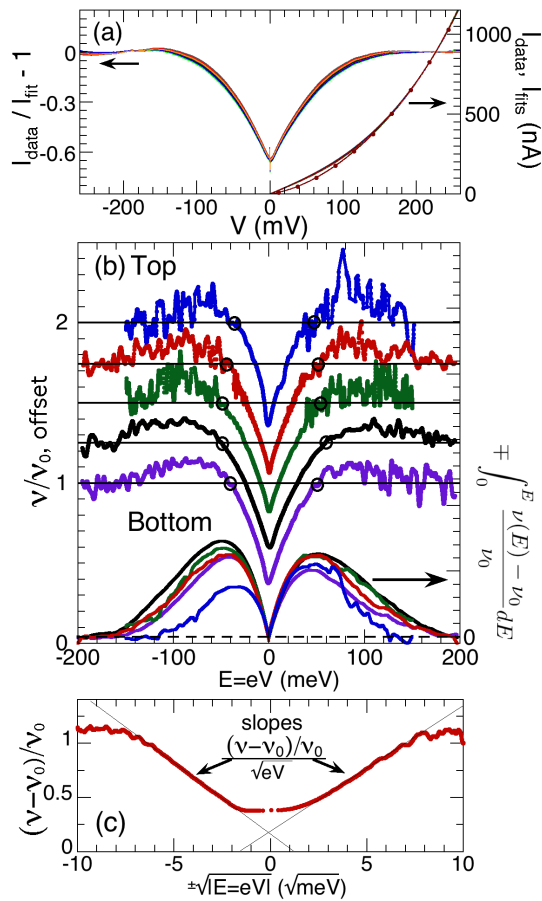}
  \caption{\label{fig:4}(Color online) Normalization procedure and results: (a)~Example of low-voltage I(V) data (circles and line) with $\mathrm{I_{fit}}$ curves (10 superimposed lines), and the corresponding  \mbox{I(V)/I$_\mathrm{fit}$(V)-1} deviation curves. Arrows assign curves to their scales. Fit intervals use a fixed upper limit 230~mV, and a lower limit varying from 100 to 200~mV. (b) Top (left axis): Set of normalized conductance curves, $\mathrm{\nu/\nu_0}$, offset by integer multiples of 0.25 for clarity. The cross-over energies, $\mathrm{E_{co}=eV_{co}}$, are marked by circles. Bottom (right axis): Integrated $\mathrm{(\nu-\nu_0)/\nu_0}$, in arbitrary units, to illustrate how the pileup of states compensates the depletion near Fermi level (dashed line denotes zero). (c) A single $\mathrm{\nu/\nu_0}$ curve plotted vs. $\mathrm{\sqrt{E=eV}}$. Slopes of the linear sections highlighted by straight lines.}
  \end{center}
\end{figure}

We have fit the $\mathrm{I_{fit}}$ to our data over a variety of voltage intervals within the range 0$-$230~mV and found that the $\mathrm{I_{fit}}$ is essentially independent of the interval choice, when the limits lie within 130$-$230~mV range. This is demonstrated in Fig.~\ref{fig:4}a, where we highlighted the difference between I(V) data and its $\mathrm{I_{fit}}$'s using the deviation curves, $\mathrm{I_{data}/I_{fit}-1}$. The spread of the deviation curves is small and not systematic with the interval.

In the top part of Fig.~\ref{fig:4}b we present a series of G(V)/G$_0$(V)=$\mathrm{\nu(E)/\nu_0}$ curves, and at the bottom of Fig.~\ref{fig:4}b we show the integral of the $(\nu-\nu_0)/\nu_0$ corresponding to the top curves. Depletion of states occurs in the rising part and pileup in the falling part of the integral curves. The curves show that states depleted near $\mu_F$ are recovered by $\sim$200~meV. However, we should consider only data below 130~meV (lower limit of fit intervals) as a reasonably reliable representation of the state recovery. The pileup of states in $\nu/\nu_0$ plots between $\mathrm{E_{co}}$ and 130~meV accounts for $\sim$80$\pm$20\% of the states depleted below $\mathrm{E_{co}}$, with the remainder presumably extending out to the band edge as anticipated from the theory. The high-voltage noise, inherent to tunneling, could lead to the apparent asymmetry in the curves, but we have not detected a systematic asymmetry in the $\mathrm{E_{co}}$ values. From data shown in Fig.~\ref{fig:4}b we determine experimentally the characteristic energy scale, $\mathrm{E_{co}}$, for the CI's in LSMO (x$=$0.36) and demonstrate that a most states pile up within a few $\mathrm{E_{co}}$, in close analogy to the well-known superconducting case.

The theories of interactions (AA) and quantum interference\cite{Abrikosov00} in quasi~2D metals in the low temperature limit, allow us to estimate the electron scattering time, $\tau$, in LSMO. The $\tau$ can be found using Eqs. (29$-$31) of Ref.~\cite{Abrikosov00} as
\begin{equation} \label{eq:tauEE}
  \tau = \hbar\left[\pi\left(\frac{\varepsilon_F\tau}{\hbar}\right)\left(\frac{\alpha\tau}{\hbar}\right) \right]^2 \left(\frac{1}{\nu_0}\frac{d\nu(\text{E})}{d\sqrt{\text{E}}} \right)^2 \hspace{0.25in},
\end{equation}
where $\alpha$ is the tight-binding coefficient for $c$-axis transport, and $\varepsilon_F$ refers to $ab$-plane Fermi energy. The last term in Eq.~(\ref{eq:tauEE}) is the square of the slopes of the linear regions of the plot $\nu/\nu_0$ vs. $\mathrm{\sqrt{E}}$, see Fig.~\ref{fig:4}c. The parameters $\varepsilon_F\tau/\hbar$ and $\alpha\tau/\hbar$ were determined by quantum interference\cite{Li00a} as $\sim$2.9 and $\sim$0.15, respectively, for the bilayered LSMO (x$=$0.40). The same experiment gave an estimate of the mean free path $\ell$$\simeq$1.4~nm. We have analyzed $\sim$20 junctions using the normalization procedure described above and Eq.~(\ref{eq:tauEE}). The resulting values of $\tau$ exhibited a Gaussian distribution, with $\tau$=16$\pm$3~fs.

Using this result we can test another, intuitive approach to CI's. If one considers the time limits on the diffusive motion of electrons in the same quasi~2D model as used above\cite{Abrikosov88}, the result for small E is 
\begin{equation} \label{eq:AAA3Denscale}
  \frac{\nu-\nu_0}{\nu_0} \approx - \int^{\hbar/|\text{E}|}_{\tau} dt \frac{\text{v}_F\lambda^2_F}{(\mathcal{D}t)^{3/2}} \propto  \sqrt{\frac{|\text{E}|}{\hbar}} - \frac{1}{\sqrt{\tau}}  \hspace{0.25in},
\end{equation}
where $\mathcal{D}$=$\sqrt[3]{\mathcal{D}^2_{ab}\mathcal{D}^{ }_c}$ is the quasi 2D electron diffusion coefficient. The crossover points $\mathrm{E_{co}}$ (see Fig.~\ref{fig:1} inset and Fig.~\ref{fig:4}b) between depleted states near $\mu_F$ and the pile-up at higher energies is defined by $\nu$=$\nu_0$. Then using $\tau$ from Eq.~(\ref{eq:tauEE}) we determine $\mathrm{E_{co}}$=$\hbar/\tau$ from Eq.~(\ref{eq:AAA3Denscale}) and compare it with the experimental values shown in Fig~\ref{fig:4}b. The AA theory above predicts $\mathrm{E_{co}}$$=$41$\pm$8~meV, which matches with the $\mathrm{E_{co}}$ values obtained from Fig.~\ref{fig:4}b, i.e., 33$-$55~meV. This verifies the intuitive formula~(\ref{eq:AAA3Denscale}).

The theory of interactions has been developed as a perturbation theory for small corrections to the DOS, in the limit of $\ell/\mathrm{\lambda_F}$$>>$1, where $\ell$=$\mathrm{v_F\tau}$ is the mean free path, and $\mathrm{\lambda_F}$ is Fermi wavelength. The DOS effect we present in Fig.~\ref{fig:4} is not so small, and the stated assumption is barely satisfied, as $\ell/\mathrm{\lambda_F}$=$\mathrm{\varepsilon_F\tau/\hbar}$$\simeq$2.9. The values of $\tau$ determined by our analysis, therefore, are not rigorously correct. Nonetheless, they are in a reasonable agreement with recent ARPES results\cite{Sun06,Mannella05}. Using $\tau$=16~fs and the $\ell$$\simeq$1.4~nm, we find a Fermi velocity, $\mathrm{v_F=\ell/\tau\sim1}$$\times$10$^7$~cm/s. This is close to the published ARPES results, $\mathrm{v_F}$=1$-$3$\times$10$^7$~cm/s. Note that ARPES experiments measure the Fermi velocity along selected directions in the $k$-space, whereas our tunneling experiment measures $\mathrm{v_F}$ averaged over the $k$-states, weighted by their contribution to tunneling.

In summary, we find that $\sim$80\% of states near $\mu_F$ depleted by Coulomb interactions are found in the range of $\sim$40$-$130~meV from $\mu_F$. This presents an analogy with superconductivity, where a large fraction of depleted states reappear within two or three times the characteristic energy. However, the characteristic energies, $\mathrm{E_{co}}$, are not universal, e.g., they depend on $\tau$ in LSMO and on the energy gap, $\Delta$, in superconductors. Our analysis yields elastic scattering time, $\tau=16\pm3$~fs, and Fermi velocity, $\mathrm{v_F=\ell/\tau\sim1}$$\times$10$^7$~cm/s, that are not too far from recently published ARPES results. We accomplished this by a consistent procedure to normalize the tunneling conductance.  This appears to be the first experiment to address an essential implication of the AA theory of CI's, the conservation of states.

\begin{acknowledgments}
The authors would like to thank Dr.~Konstantin Matveev for valuable consultations on the theory of Coulomb interactions. This research was supported by the US Department of Energy, Basic Energy Sciences$-$Materials Sciences under Contract No.~DE-AC02-06CH11357 at the Argonne National Laboratory operated by UChicago Argonne, LLC.
\end{acknowledgments}

%\bibliography{BiblioErange}

\begin{thebibliography}{25}
\bibitem[{fno({\natexlab{c}})}]{fnote2}
\bibinfo{note}{The characteristic energy scale is $\hbar/\tau$$\sim$100 meV
  (using ARPES value of $\tau$). If state conservation occurs above 230~meV,
  the low-voltage parabolic fit of data would not go through the origin. That
  option gives a value of $\nu(0)/\nu_0\simeq$~0.1 that is far too small to be
  reconciled with the mild low-temperature decrease of the bulk conductivity
  (Fig.~\ref{fig:1}).}
\end{thebibliography}

\end{document}